# Effect of pH on photocatalytic degradation of Methylene Blue in water by facile hydrothermally grown $TiO_2$ Nanoparticles under Natural Sunlight

Uttama Kumar Saint[1], Suresh Chandra Baral[1], Dilip Sasmal[1], P. Maneesha[1], Sayak Datta[1], Farzana Naushin[2] and Somaditya Sen[1*]

[1]*Department of Physics, Indian Institute of Technology Indore, Indore, 453552, India*

[2]*School of Biotechnology, Jawaharlal Nehru University, New Delhi-110067, India*

* *Corresponding author:* sens@iiti.ac.in

**Abstract:**

Each year, the production of synthetic dye wastewater reaches a trillion tons, posing a significant challenge to addressing water scarcity on a global level. Hence, the treatment of wastewater to prevent water scarcity is of prime importance, and failing to do so will increase ecotoxicological risks and human health. Textile wastewater contains harmful dye. Photocatalytic degradation of such dye-contaminated wastewater is crucial to purifying the dye-contaminated water. However, this process takes time, uses high-power lamps, and is expensive. Here, we report the effect of the concentration of precursor on the size and surface morphology of $TiO_2$ nanostructures prepared by facile hydrothermal synthesis and its ability to perform as a photocatalyst to degrade the most common industrial textile dye, methylene blue (MB), under natural sunlight. The impact of particle size on the photocatalytic activity and photocarrier migration rate was thoroughly examined. Also, the effect of pH on adsorption and photocatalytic degradation has been evaluated in detail. With several optimized conditions, almost complete dye degradation was achieved within 40 minutes under the direct illumination of natural sunlight. The enhanced photocatalytic performance can be correlated to the synergetic effect of a higher charge transfer mechanism, good catalytic active surface area availability (386 $m^2/g$), and several optimized parameters that affect the reaction efficacy. Additionally, repeated use of NPs without sacrificing performance five times confirmed its stability and Sustainability as a promising candidate for large-scale industrial textile wastewater remedies.

## 1. Introduction:

Water pollution has become one of the most critical worldwide concerns of the twenty-first century. One of the significant sources of water pollution is untreated wastewater from textile industries. The synthetic azo dyes, anthraquinone, and sulfide in this wastewater have the potential to harm human and animal life in the long run due to their cancer-causing, birth defect-causing, and genetic mutation-causing properties. [1]. Therefore, the removal of organic pollutants from the contaminated water is of paramount importance. There have been numerous methods to treat this wastewater before sending it to the environment. Among these, biological treatment processes, chemical precipitation, adsorption, chemical oxidation, and electrochemical oxidation, have been proven to be much more effective [2]. However, the complete mineralization of this wastewater still needs to be improved with the single traditional treatment method. The photocatalysis degradation process is one of the emerging technologies to treat wastewater. It has the potential to degrade the dye and mineralize the wastewater completely. In addition, the demand for photocatalysis has surged due to its moderate reaction, minimal secondary pollution, and potential for high reusability[3]–[14].

Recently, semiconductor materials have been discovered to be effective photocatalysts for decomposing organic toxins. Among these, $TiO_2$ holds significant promise as a photocatalyst due to its cost-effectiveness, non-toxicity, high chemical stability, and mass-production capabilities [15], [16]. However, $TiO_2$'s rapid photo-generated charge recombination significantly compromises its photocatalytic efficiency [17]. The decreasing size and complex morphology of the catalyst increase its surface, thereby increasing the surface-to-volume (S/V) ratio. This increases the reactive area available for the interaction of the catalyst with the dye molecules and results in better catalytic activity [18]. Moreover, the increased S/V ratio also facilitates internal electron/hole pair formation and recombination. However, two major negative effects are associated with decreased particle size of $TiO_2$ nanoparticles (NPs). Firstly, the surface recombination of electrons and holes will increase, and secondly, it lowers the utilization of photons [19]. Therefore, an optimum particle size may occur where an equilibrium between these two opposite effects can be reached [20]. Unlike bulk $TiO_2$, granular $TiO_2$ NPs have a large surface area and a broadened band gap due to quantum confinements. The facile hydrothermal synthesis route is popular for generating a controlled morphology. It requires relatively mild operating conditions (reaction temperatures < 300 °C). It is an environmentally friendly, one-step synthesis procedure and produces NPs with good dispersion in solution. The surface morphology can be varied by varying the reaction temperature, time, and pressure. The pH of the reaction medium also controls the NP's size and morphology [21]. However, reports are contradictory, claiming opposite trends: increasing pH leads to increasing particle size [22], [23], and also decreasing particle size [24]. The latter also reported reduced particle aggregation. Hence, the effect of pH on particle size is still unclear.

Conversely, photocatalytic activities can be influenced by several elements such as the wavelength and intensity of light, the reactants and their concentrations, the amount of catalyst, and the temperature of the reaction medium [25]. The pH of the medium significantly influences the degradation of dyes in photocatalytic reactions. This is because the three potential methods for dye degradation, including hydroxyl radical attack, positive hole oxidation, and electron reduction in the conducting band, are all impacted by pH in varying ways. [26]. The level of contribution from each parameter is determined by the substrate's nature and the pH. The pH of the solution alters the electrical double layer at the solid electrolyte interface, which influences the adsorption-desorption processes and the effective separation of electron-hole pairs on the semiconductor surface. Because $TiO_2$ is amphoteric, its surface can carry either a positive or negative charge[27]. So, by varying the pH of the solution, one can influence the adsorption of dye molecules onto the $TiO_2$ surfaces and, thus, the photocatalytic degradation [28]. The dark adsorption of dyes on the $TiO_2$ catalyst was measured to be increased from 1.43% at pH 3 to 24.34% at pH 5.8 and then to 61.47% at pH 9.95, respectively. Subsequently, photodegradation percentages under UV light for pH 3, pH 5.8, and pH 9.95 were 65%, 46%, and 96%, respectively [29]. Similar results were also reported, where an increase in the percentage of degradation occurred under sunlight exposure by changing the pH level from 3 to 11, with the percentages escalating from 53 to 86 % [30].

Generally speaking, a high-power UV/Xenon lamp (250~300W) is utilized in these experiments through an expensive apparatus. These lamps are expensive by themselves and consume considerable power, affecting the environment by emitting an equivalent amount of $CO_2$ [31], [32]. Hence, solar light-driven photocatalytic oxidation (SPO) is a cost-effective and more environmentally friendly approach to textile wastewater treatment.

Here, we report the effect of the concentration of precursor on the size and surface morphology of $TiO_2$ nanostructures prepared by facile hydrothermal synthesis and its ability to perform as a photocatalyst to degrade the most common industrial textile dye, methylene blue (MB), under natural sunlight. The impact of particle size on the photocatalytic activity and photocarrier migration rate was thoroughly examined. Also, the effect of pH on adsorption and photocatalytic degradation has been evaluated in detail. The results highlight the significant role of particle size in modulating photocatalytic structure and efficacy, and the reaction pH, which controls the surface charge, affects the same.

## 2. Experimental Section

### 2.1. Materials and synthesis:

The chemical used here for the preparation was purified and of good quality. Titanium tetra-isopropoxide (TTIP) (97%, Sigma Aldrich), Ethanol (99%, Sigma Aldrich), Methylene Blue dehydrate (99.9%, alfa aeser), Sodium Hydroxide (NaOH, 99.9%, alfa aeser), and De-ionized water (DIW, 18.2 Mega-Ohm-cm).

$TiO_2$ nanoparticles (NPs) were synthesized using an autoclave-facilitated hydrothermal process from TTIP and ethanol solution. TTIP decomposes and precipitates in the open air almost instantaneously. Hence, the use of TTIP in a controlled synthesis of $TiO_2$ nanoparticles requires hydrolysis of the TTIP in ethanol to finetune the formation at a desired rate. An appropriate amount of 6 mM of TTIP was taken in a beaker and was immediately hydrolyzed using 100 ml of ethanol to avoid precipitation while stirring continuously at 300 rpm. The pH of the final solution was found to be ~ 4.6. After 3 hours of continuous stirring, the solution was transferred to 150 ml of Teflon autoclave and sealed. The autoclave was maintained at a temperature of 120 ℃ for 24 hours in a thermal chamber, followed by natural cooling to room temperature. The solution contained white dissolved nanopowders. The solution was centrifuged at 5000 rpm to extract the nanostructures and washed several times with DI water to eliminate any other impurities. The final sample was dried using a thermal chamber at 80 ℃ for 12 hours, followed by annealing at 450 ℃ for 2 hours. The sample was then ground. This sample was named "S1". A second sample ("S2") was prepared in similar steps with ten times the concentration of TTIP, i.e., 60 mM of TTIP. For this solution, the pH was 5, i.e., less acidic than the previous one. The rest of the process was similar. A schematic representation of the synthesis process of $TiO_2$ nanostructure is given in Figure 1. Both samples, S1 and S2, were used for further characterization.

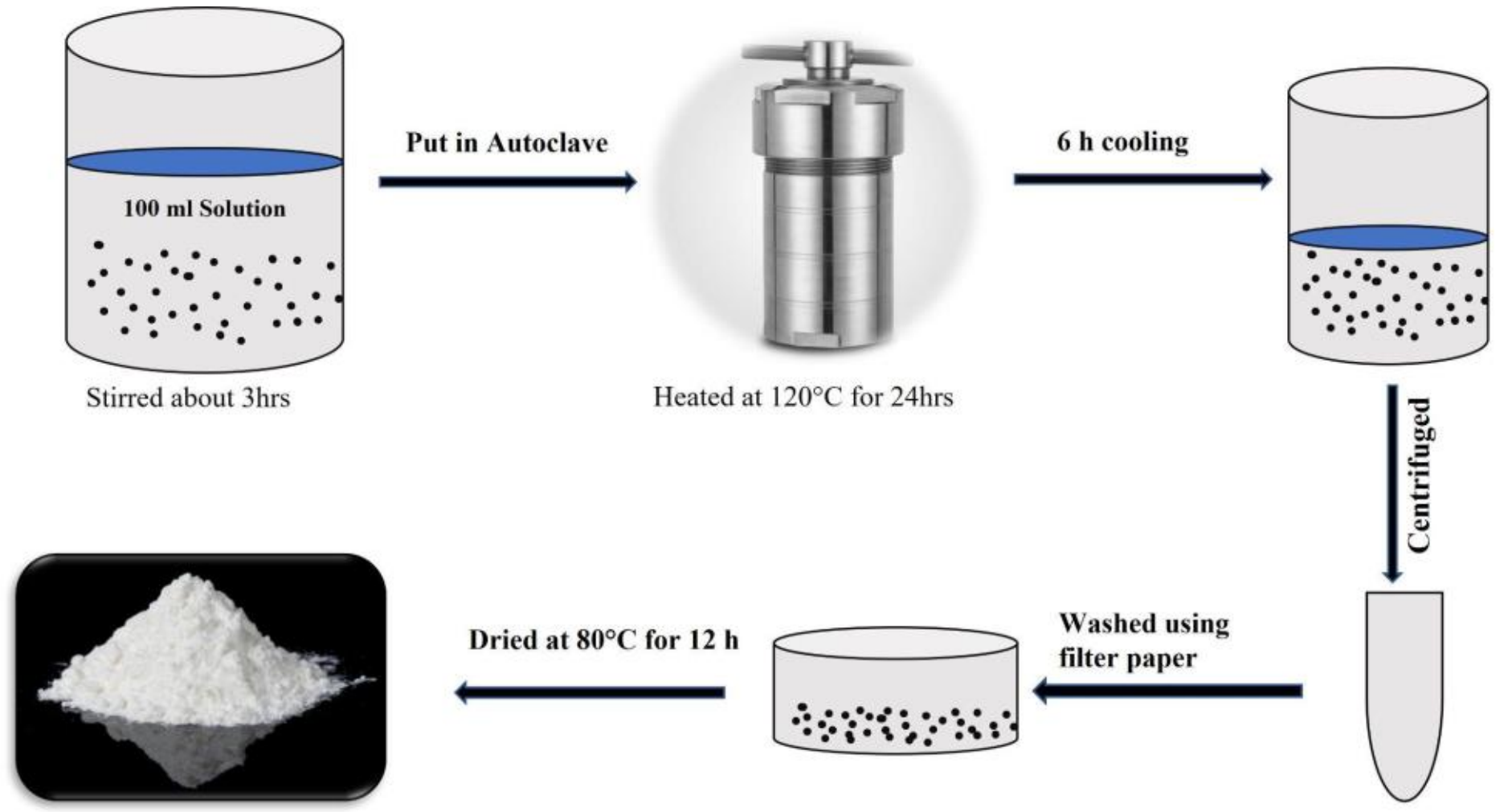

***Figure 1****: Schematic representation of the synthesis process of $TiO_2$ nanostructure.*

The surface morphology of both samples was evaluated utilizing Field Emission Scanning Electron Microscopy (FESEM, JEOL, JSM-7610 F). The specific surface area and pore volume were determined using a Quantachrome Autosorb iQ2 BET Surface Area and Pore Volume Analyzer, while the structural characteristics were identified through X-ray diffraction (XRD) analysis carried out with a Bruker D2 Phaser X-ray diffractometer with Cu-Kα radiation (λ= 1.54 Å). Using Fullprof software, the lattice parameters, bond lengths, and bond angles were calculated using Rietveld refinement. The HORIBA Scientific Lab-RAM HR Evolution Raman Spectrometer with a laser light source operating at 633 nm and power output over 300 mW was employed to determine the phonon modes of the $TiO_2$ lattice at room temperature. The optical bandgap was measured using a Research India spectrometer and the DRS method. The pore size distribution and specific surface area were estimated from $N_2$ adsorption-desorption measurements analysed with BET surface area analysis.

Under direct sunlight at the IIT Indore campus in India, all samples were tested for their photocatalytic performance, with a light intensity of 65000 lx ≈ 474 $Wm^{-2}$ (2 times smaller than AM 1.5G Sun) and a temperature of around 35°C. A stock solution of pollutant dyes was prepared by dissolving 10 mg of the dyes in 1 liter of deionized water, resulting in a concentration of 10 ppm. The dye methyl blue (MB) was utilized for this experiment. A sample of around 20 mg was combined with 20 ml of each dye solution in a beaker to achieve a catalyst concentration of roughly 1 mg/mL. Similar concentrations of catalyst 0.25 mg/mL, 0.5 mg/mL, 0.75 mg/mL, 1 mg/mL, and 1.25 mg/mL were also tested. The mixed solutions were stirred continuously for 60 minutes in the dark to ensure effective dispersion and adsorption equilibrium. Variations were monitored by analyzing absorption data using a Research India UV-visible spectrophotometer. A 3 ml solution was taken out and properly centrifuged after each stage of the photocatalytic degradation process. The absorption spectra of this extracted solution were studied. The photocatalytic degradation efficiency of the organic dyes was calculated using the following formula: $Degradation\ (\%) = [(A_0 - A_t)/A_0] * 100$ ; where, $A_0$ is the absorbance of dye solution before the photo-irradiation, $A_t$ is the absorbance of solutions after photo-irradiation for a specific time t. To test the pH dependence photocatalytic activity, a 0.01M solution of NaOH was utilized to vary the pH of each dye solution, with pH levels being monitored closely using a high-precision pH meter (Labtronics pH Meter).

## Results and Discussions

The morphology of the synthesized nanostructure was analyzed using FESEM images [Figure 2], which revealed highly agglomerated spherical NPs for both samples. The particle size was estimated by using Image J software [33]. The NPs' average size was 14 nm for S1 and 15.5 nm for S2, with a standard deviation (σ) of 3.81 nm and 4.17 nm, respectively.

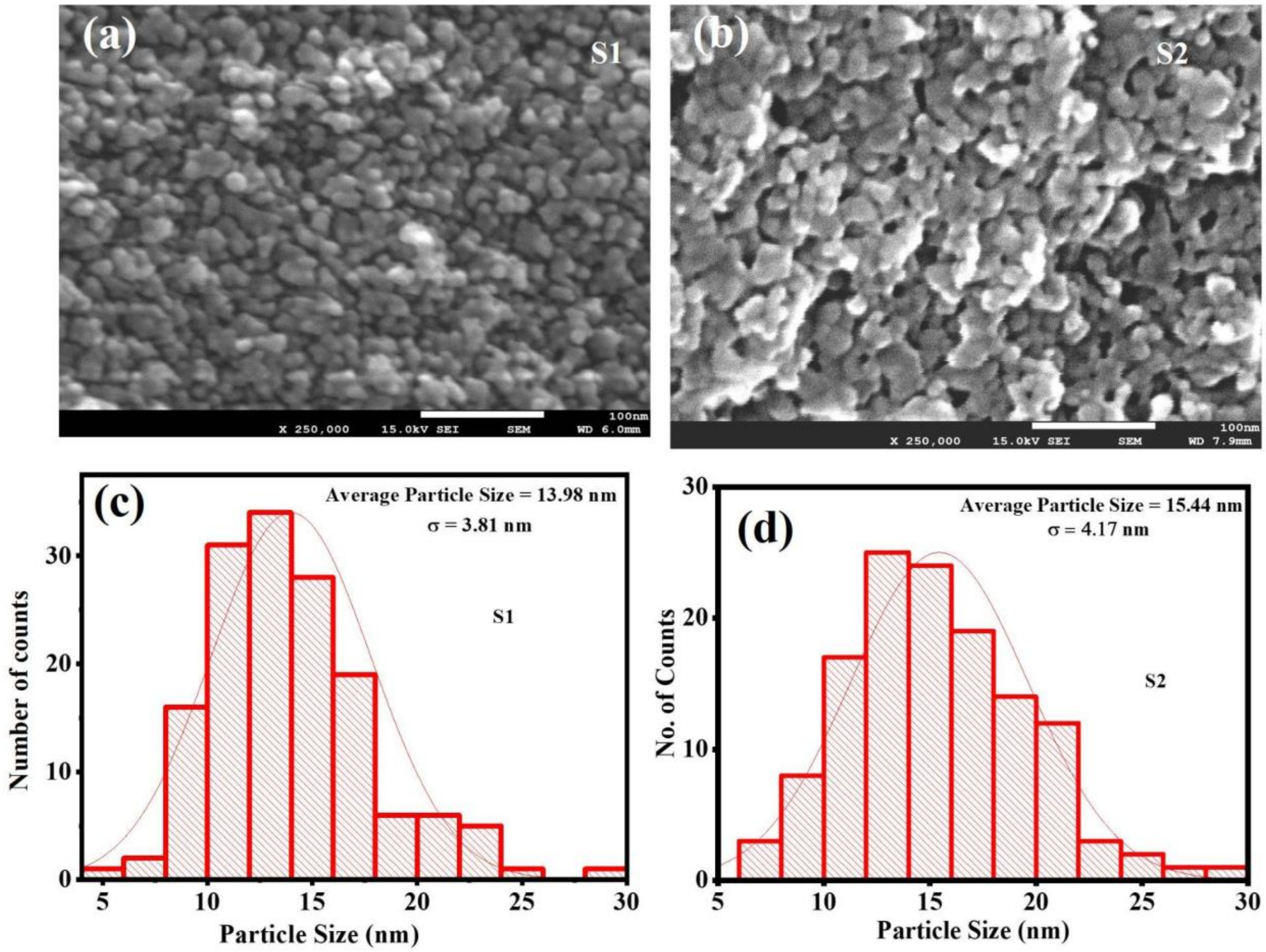

***Figure 2***. *The surface morphology of the prepared samples was observed from FESEM (a, b), while the average particle size was estimated using Image J software (c, d) for S1 and S2, respectively.*

Distinct hysteresis loops of type IV isotherm were observed from the BET data, indicating a mesoporous nature for both samples [34]. The specific surface area was 386 $m^2/g$ for S1 and 102 $m^2/g$ for S2. Hence, the S1 sample had a larger interactive surface area than the S2 sample. Accordingly, the total pore volume was 0.47 cc/g for S1 and 0.185 cc/g for S2. The pore size distributions [Figure 3(c, d)] were similar for both samples: 3 to 427 nm for S1 and 3.4 to 461.69 nm for S2. On the other hand, the average pore size distribution was 4.878 nm for S1 and 7.815 nm for S2, confirming the mesoporous structure.

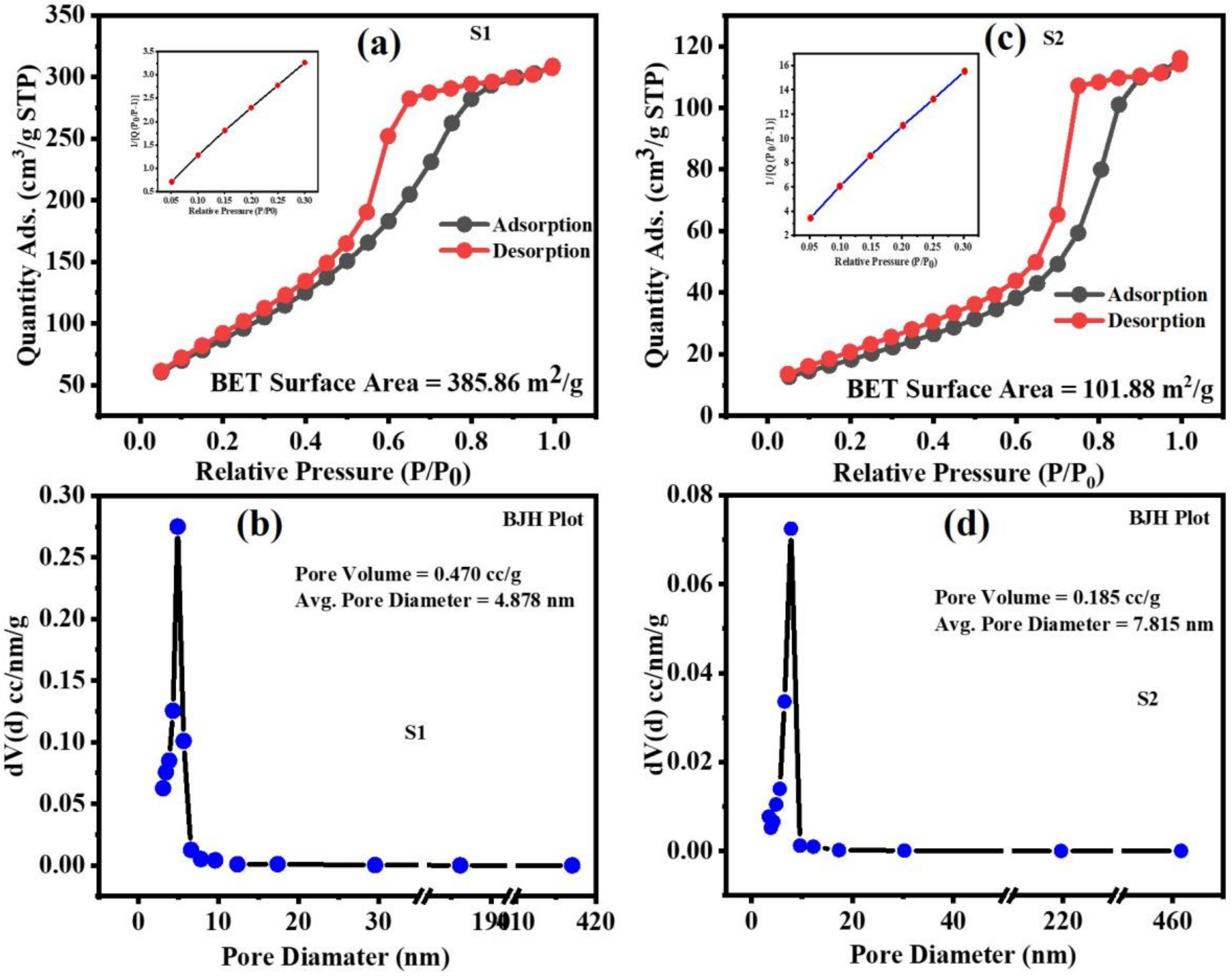

***Figure 3:*** *(Upper panel) specific surface area calculation from $N_2$ adsorption-desorption isotherms, while the insets represent the corresponding BET plots (a, c), (Lower panel) Pore size distribution, was confirmed by BJH plots considering the desorption data (b, d) for S1 and S2 respectively.*

XRD confirmed the phase purity after the surface morphology and specific surface area determination. The samples are of mixed phases of dominant anatase (JCPDS 021-1272; $D_{4h}$ space group) and impurity rutile (JCPDS 021-1276; *P42/mnm* space group) phases. XRD pattern of S1 sample [Fig. 4(a)] exhibits seven peaks belonging to $TiO_2$ anatase phase (101) at 25.53$^O$, (004) at 38.14$^O$, (200) at 48.17$^O$, (105) at 54.81$^O$, (204) at 63$^O$, (116) at 69.73$^O$, and (215) at 75.43$^O$. Similarly, the S2 sample exhibits eleven peaks (101) at 25.99$^O$, (004) at 38.28$^O$, (112) at 39.13$^O$, (200) at 48.45$^O$, (105) at 54.25$^O$, (211) at 55.45$^O$, (204) at 63.02$^O$, (116) at 68.98$^O$, (220) at 70.42$^O$, (215) at 75.19$^O$, and (301) at 76.15$^O$. Hence, all the peaks have been shifted towards higher angles, indicating a contraction of the d-spacings in the S2 sample. On the other hand, one rutile (110) reflection peak is found at 30.91° and 31.64° for S1 and S2, respectively, revealing an expansion of the (110) rutile d-spacing. The crystallite size and microstrain were calculated using the Williamson-Hall method. Here, crystallite size and microstrain have been calculated by assuming the Uniform Deformation Model (UDM), which considers a consistent isotropic strain in all crystallographic directions, thereby assuming homogeneous material properties in all measurement directions. The WH equation is given by: $\beta \boldsymbol{Cos\theta = 4\varepsilon Sin\theta} + \frac{\boldsymbol{K\lambda}}{\boldsymbol{D}}$**;** where $\boldsymbol{\beta}$ = FWHM of the peak, $\boldsymbol{\Theta}$ = peak position, ε = Strain, K = shape factor which has taken as 0.94 (assumed as spherical shape), $\boldsymbol{\lambda}$ = Wavelength of incident Cu-k$_{\boldsymbol{\alpha}}$ x-ray = 1.54Å and D = crystallite size. A linear fit has been done between $\beta Cos\boldsymbol{\theta}$ *and* 4Sin$\boldsymbol{\theta}$ with the goodness of fit ($\boldsymbol{\chi}^2$) 0.968 for S1 and 0.996 for S2. The higher pH value of the precursor solution seems to help reduce the lattice micro-strains from 7.23 for S1 to 1.84 for S2, which increases the crystallite size from 7.93 nm for S1 to 11.53 nm for S2.

The dominant $D_{4h}$ anatase $TiO_2$ phase has a primitive unit cell with six atoms (two Ti and four O atoms). The six Raman active phonon modes are $A_{1g}$, two $B_{1g}$, and three $E_g$ modes: $A_{1g}$ + $2B_{1g}$ + $3E_g$ [35]. These modes are associated with the first-order Raman spectrum where three $E_g$ modes are due to the symmetric vibration of O-Ti-O, two $B_{1g}$ modes are associated with the symmetric bending vibration of O-Ti-O, and one $A_{1g}$ mode due to the antisymmetric bending vibration of O-Ti-O. S1 and S2 have four significant peaks, ascribed to $E_g$, $B_{1g}$, $B_{1g}$+$A_{1g}$, and $E_g$ phonon modes. The S1 sample reveals these modes at ~148.6 cm$^{-1}$, 400.3 cm$^{-1}$, 515.6 cm$^{1}$, and 637.5 cm$^{-1}$. A minor peak centered at 198.8 cm$^{-1}$ belonging to the $E_g$ mode was also visible. S2 revealed the same combination at 143.4 cm$^{-1}$, 394.7 cm$^{-1}$, 514.4 cm$^{-1}$, and 636.9 cm$^{-1}$ and a minor peak centered at 195.9 cm$^{-1}$. The FWHM of the most intense $E_g$ mode, centered, seemed to sharpen for the S2 more than the S1 from 21.5 cm$^{-1}$ to 10.5 cm$^{-1}$, hinting at a better crystalline quality of the S2 sample.

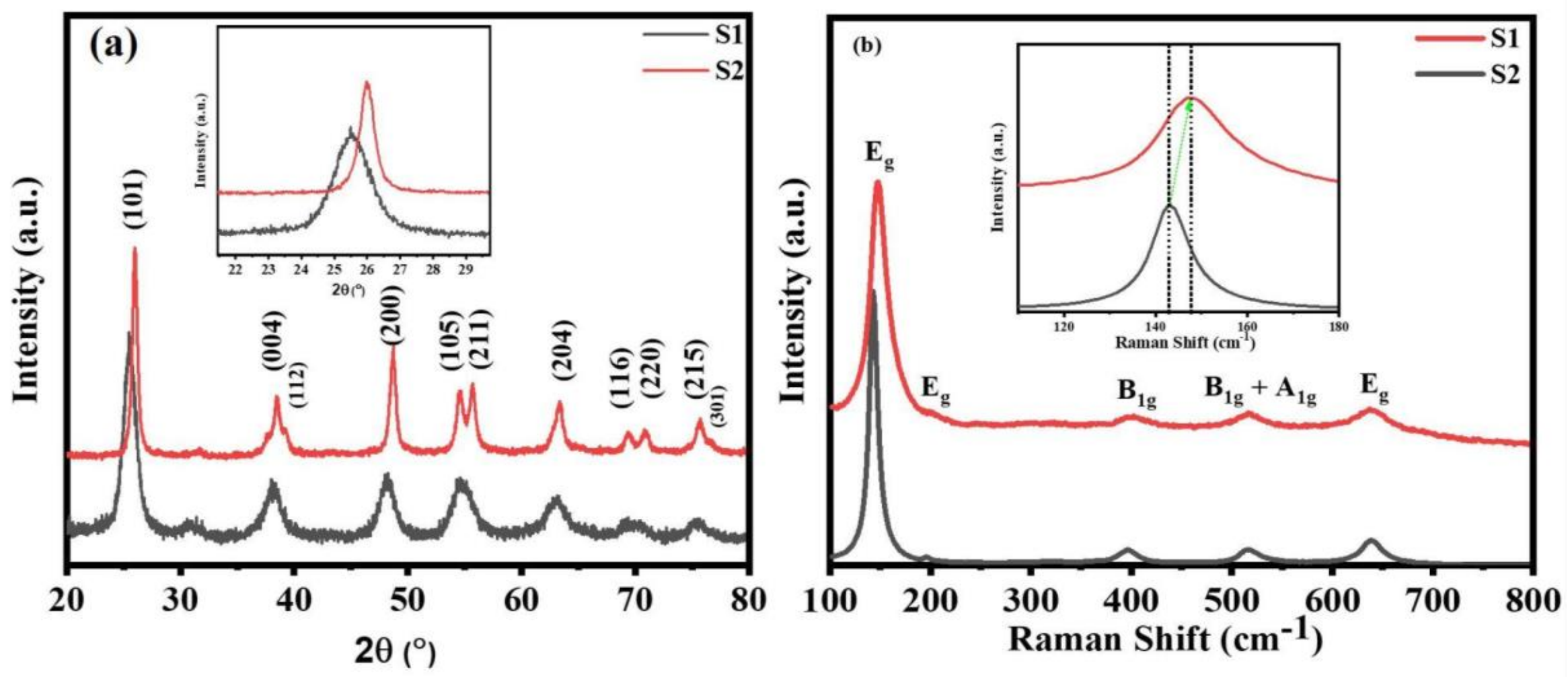

***Figure 4**: The phase purity was confirmed using (a) XRD pattern and (b) Raman spectra for S1 and S2, respectively. The insets in Figure (a) show the XRD peak broadening and shifting towards lower 2θ values, while the insets in Figure (b) show the broadening and shifting of Raman peaks towards lower wavelengths, confirming the small particle size of the S1 sample compared to S2 in both cases.*

The anatase $TiO_2$ is known to have an indirect bandgap of 3.2 eV [36], while the rutile phase is known to have a direct bandgap of 3.0 eV [37]. However, there are also reports of deviation from the above observations [38]. As the powder is opaque, the optical band gap was assessed using the DRS data of the materials. An absorption coefficient, $\alpha$, was evaluated from the equivalent, Kubelka-Munk function, $F(R) = (1\text{-}R)^2/2R$ [39], where $R$ is the reflectance of the materials. The value of $\alpha$ was evaluated from the Tauc equation: $(\alpha h\nu)^{1/n} = A(h\nu\text{-}E_g)$, where ν is the frequency of photon, h is Planck's constant, a proportionality constant (A), and the bandgap ($E_g$) for both direct allowed transitions (n = 1/2) and indirect transitions (n = 2) between the conduction band (CB) and valence band (VB). The optical band gap, $E_g$, was calculated from the slopes of the Tauc plots ($(\alpha h\nu)^2$ *versus* $h\nu$ plots) for both samples [40] [Figure 5]. The $E_g$ was found to be 3.33 eV for S1 and 3.23 eV for S2 samples. Lattice disorder and other defects can create energy levels inside the band gap. These states can be near the conduction or the valence band edges or at different energy values within the forbidden gap. The states near the band edges are generally a result of the lattice disorder and are responsible for the tailing of the band edges. These tailing energy states are referred to as Urbach tail states and correspond to the energy Stored in these states, which is the Urbach energy, $E_U$. The higher the disorder in the lattice, the higher the value of the $E_U$ of the materials. The tailing of the bands results in an exponential growth of $\alpha$ near the band edge. This exponential nature near the band edge can be fitted to obtain the amount of energy stored in the disordered lattice from the equation $\alpha = \alpha_0 exp(E/E_u)$. The $E_U$ was estimated from the linear fitting of the linear parts of the $ln[\alpha] = ln[F(R)]$ versus $h\nu$ plots [41]. The $E_U$ of S1 was estimated as 104.9 meV, higher than 88.12 meV for S2. This indicated a more significant disorder in the S1 sample than in S2.

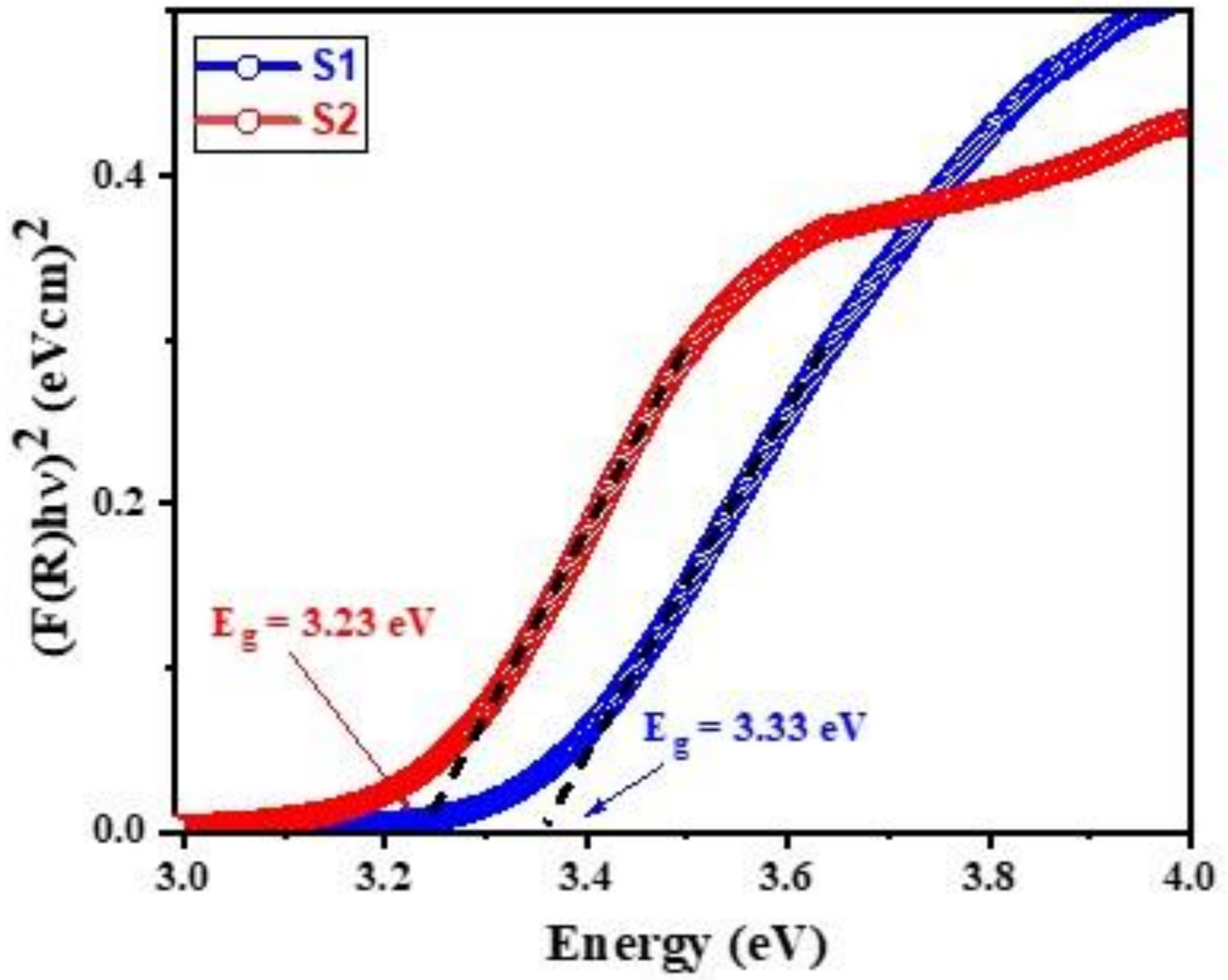

***Figure 5:*** *Tauc Plot showing $E_g$ values of 3.33 eV and 3.23 eV for S1 and S2, respectively.*

## 3. Photocatalytic degradation of Methylene Blue

Generally, photocatalytic activity experiments require a costly apparatus using a high-power (250~300W) UV/Xenon lamp. The lamp is both costly and requires a considerable amount of electricity. Such high power consumption has an equivalent carbon dioxide emission, which affects the environment [31]. Hence, in this study, pure sunlight was used to study the effect of the same on photocatalytic degradation of MB dye. The activity of both samples as catalysts was carefully verified. Note that the pH of the initial MB solution was measured to be ~ 6.5, as mentioned by existing literature [24]. The degradation of the dye solution is nearly non-existent through direct photolysis, even when catalysts are not present. Under sunlight exposure, the MB dye experiences a significant degradation following a first-order pattern. The rate constants of the photocatalytic degradation kinetics were 0.053 $min^{-1}$ for S1 and 0.041 $min^{-1}$ for S2, respectively [Figure 6]. The dark adsorption of the dye molecules on the surface of the samples at time $t = 0$ was 20% for S1 and 12% for S2, while the Photocatalytic degradation percentage was 99% for S1 and 96% for S2 under 80 min irradiation of direct sunlight.

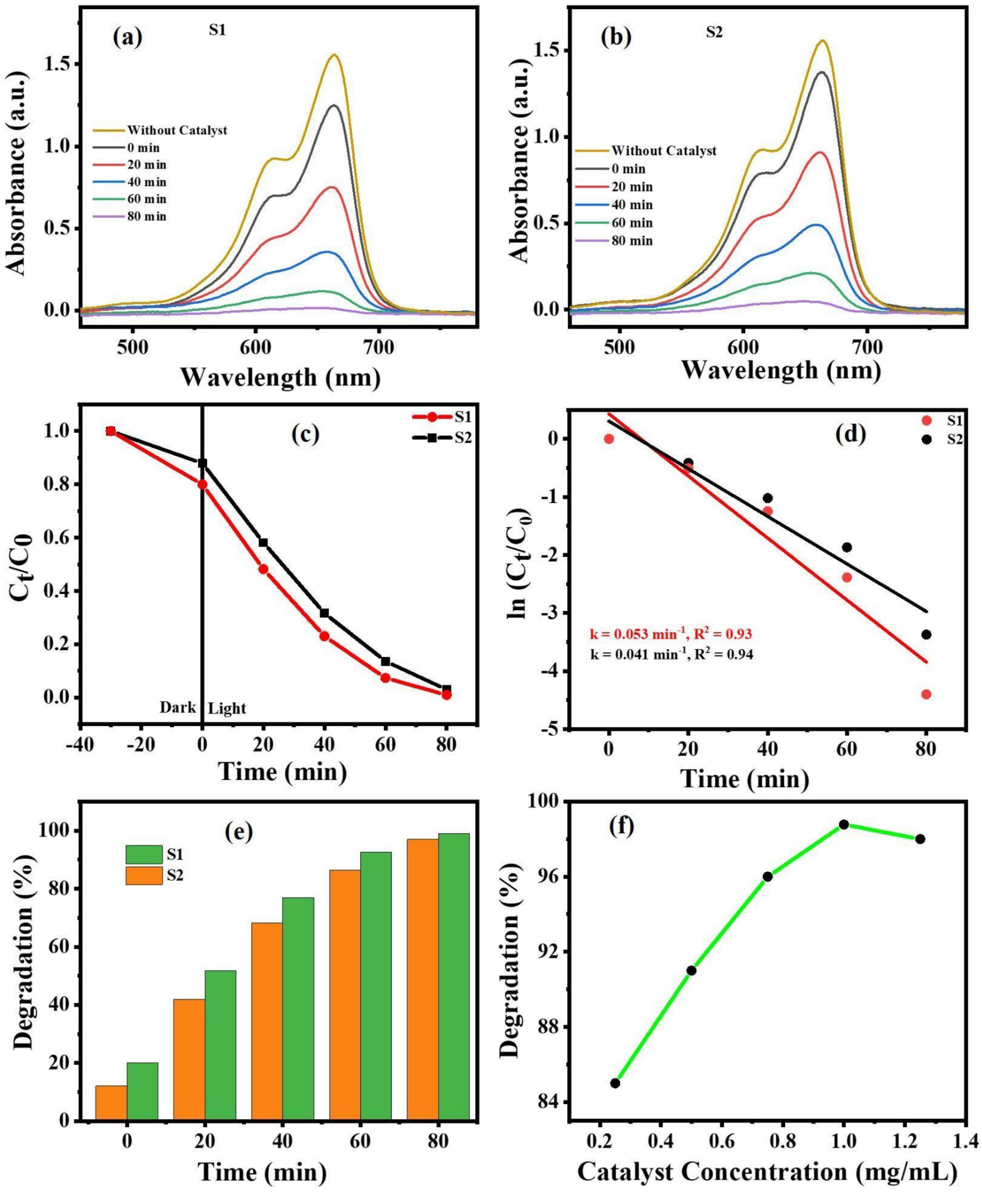

*Figure 6: Photocatalytic properties of the $TiO_2$ NPs (with catalyst concentration = 1 mg/mL) revealed from (a, b) the reduction of absorbance, (c) Relative concentration change (d) Rate of Change of concentration (e) Degradation percentage of with light-irradiation time, while the effect of catalyst concentration can be observed from (f) showing catalyst concentration of 1mg/mL gives*

*a better degradation percentage, taking the S1 sample as a model catalyst for 80 minutes under direct sunlight irradiation.*

### 3.1. Effect of Catalyst Concentration

The concentration of the photocatalyst determines the rate of the photocatalytic process. The photocatalytic degradation was verified for different concentrations: 0.25 mg/mL, 0.5 mg/mL, 0.75 mg/mL, 1 mg/mL, and 1.25 mg/mL. The total time of photo radiation was kept for 80 minutes [Figure 6(f)]. Increasing the catalyst loading from 0.25 mg/mL to 1 mg/mL leads to a rise in MB degradation from 85% to 99%, followed by a decrease to 98% for 1.25 mg/mL concentration. The increase in photocatalytic degradation with an increase in the concentration of the photocatalyst can be due to the increase in the number of active sites on the photocatalyst surface, which also increases the formation of OH• radicals. The ideal concentration for the catalyst in the solution is 1 mg/mL, as surpassing this amount can cause the solution to become turbid and impede the reaction, leading to a decrease in degradation percentage [42].

### 3.2. Effect of pH of the reaction medium

Adsorption of the dye molecules on the surface of the photocatalyst plays a vital role in the effective degradation and mineralization of the organic dyes in water. Thus, the effect of pH on the photocatalytic degradation of methylene blue was investigated for both samples. The adsorption of the dye molecule on the surface of the catalyst increased with increasing the pH of the solution [Table 1]. This can be attributed to the fact that as the pH of the solution increases, the number of $OH^-$ ions inside the solution increases. Hence, the $TiO_2$ surface acquires more negative charge adsorbing the cationic type MB dye [43]. The adsorption process can be an electrostatic interaction between the cationic dyes and the photocatalyst, which increases with increasing pH.

***Table 1.*** *Equilibrium adsorption percentage of methylene blue on the photocatalyst surface at different pH.*

| Sample Name | pH 8 | pH 9 | pH 10 | pH 11 |
|---|---|---|---|---|
| **S1** | 52.8 % | 55.65 % | 73.01 % | 69.36 % |
| **S2** | 41.20 % | 52.01 % | 70.74 % | 61.38 % |

After calculating the equilibrium adsorption percentage of MB on the photocatalyst surface at different pH, the solution was illuminated by the natural sunlight, and the corresponding absorbance of the photo-irradiated solution was recorded for 40 min [Figure 7, 8]. A first-order type transition is observed with photo-irradiation. The rate constants and degradation percentage for both samples [Figure 9] were determined and summarized in Table 2 for different pH values.

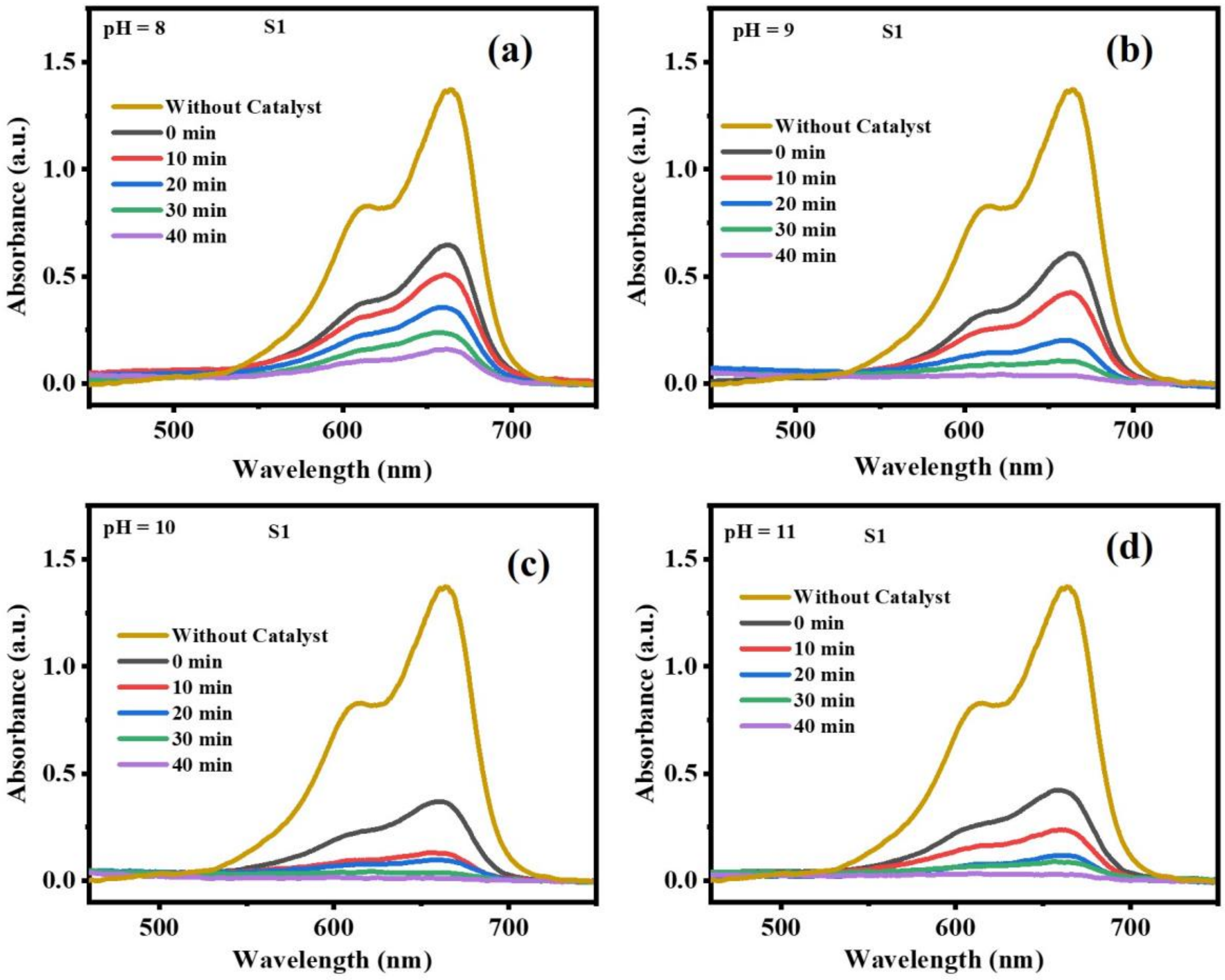

***Figure 7:*** *(a, b, c, d) shows the absorbance vs Wavelength curve of the S1 sample at pH 8, 9, 10, and 11 under sunlight.*

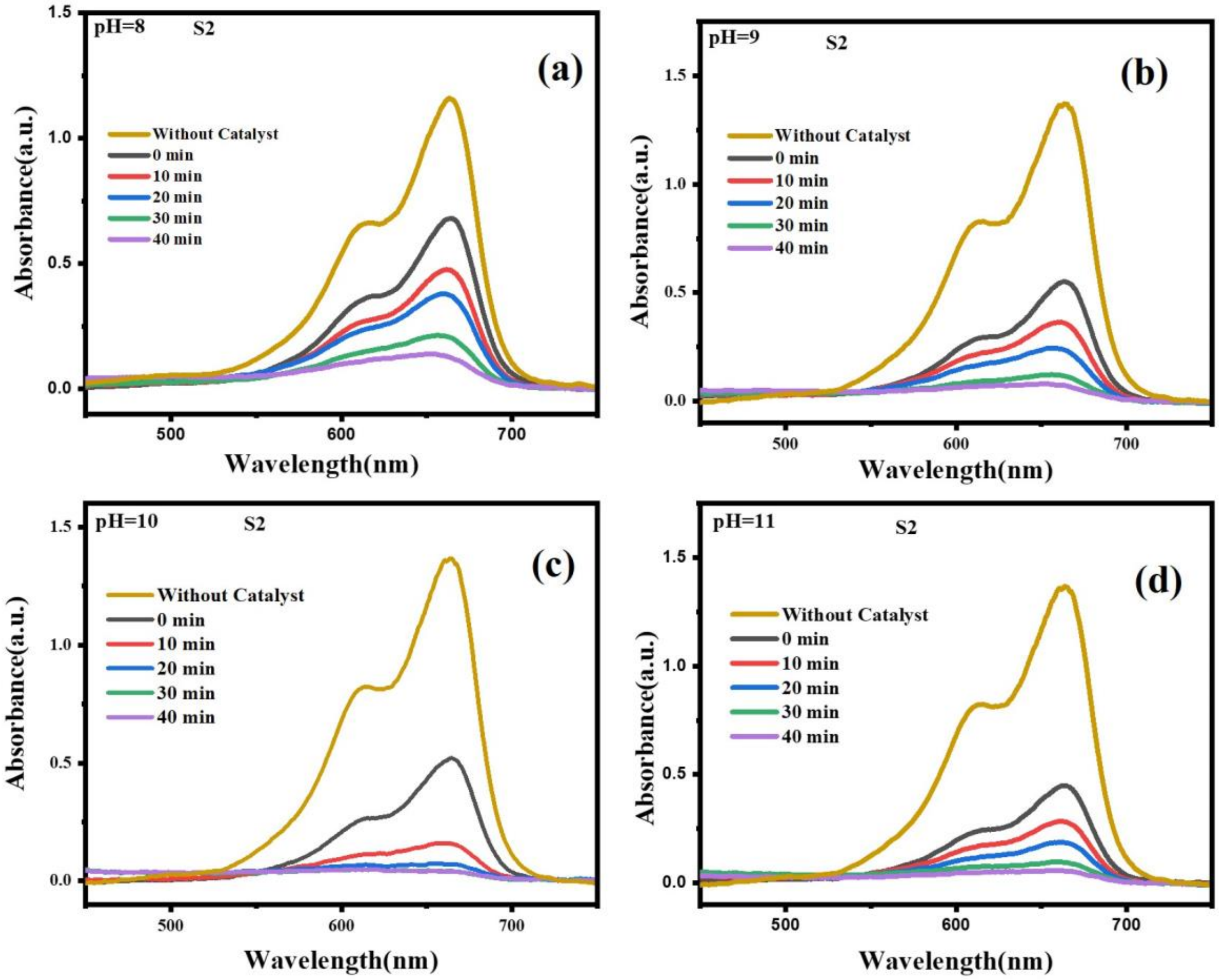

***Figure 8:*** *(a, b, c, d) shows the absorbance vs Wavelength curve of the S2 sample at pH 8, 9, 10, and 11 under sunlight.*

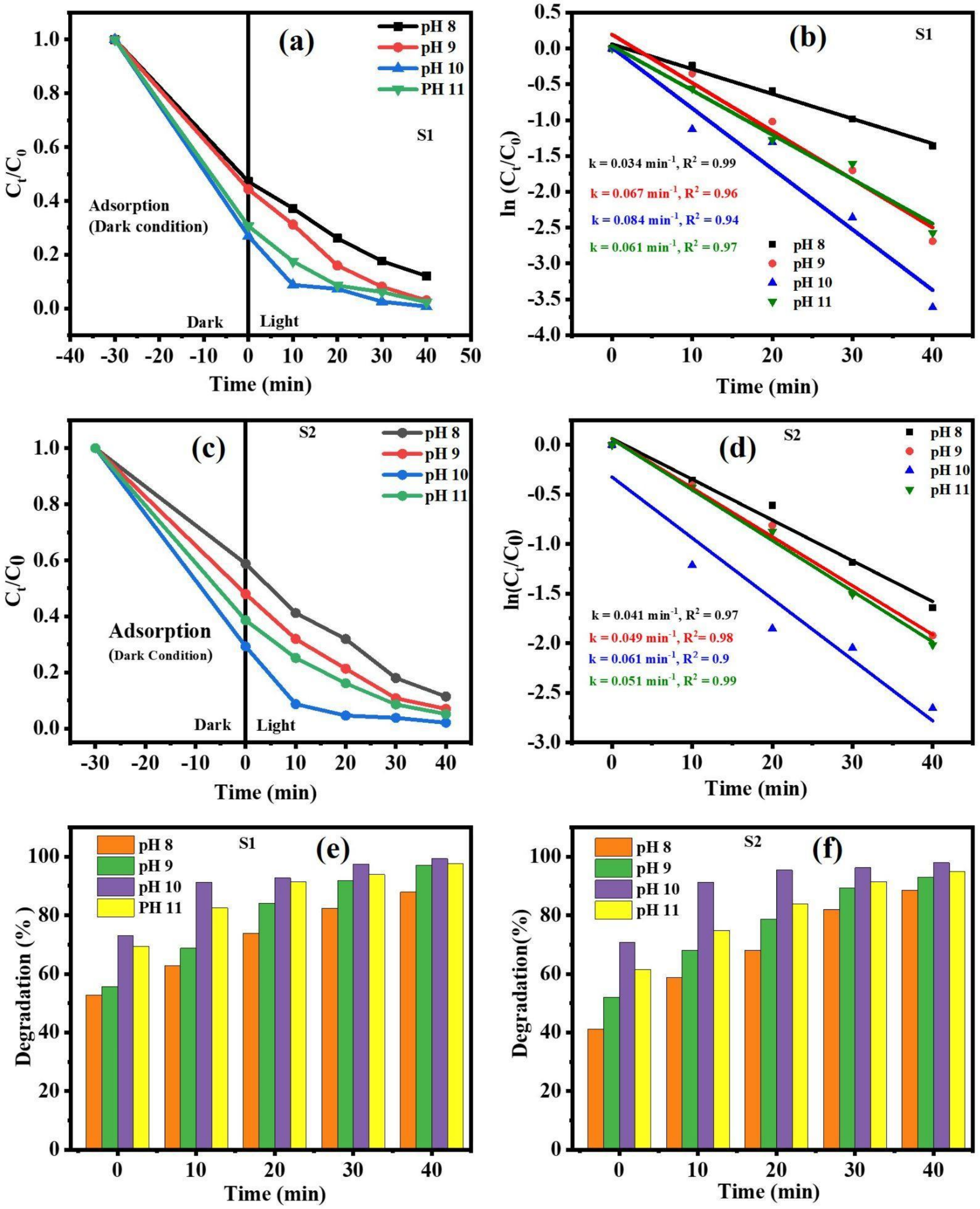

***Figure 9:*** *(a, c) relative concentration change, (b, d) calculated rate constants, and (e, f) degradation percentage for both S1 and S2 samples, respectively.*

***Table 2.*** *Equilibrium degradation percentage and rate constant of MB on the photocatalyst surface at different pH levels.*

| **pH** | **S1** | | | **S2** | | |
|---|---|---|---|---|---|---|
| | **Degradation (%)** | **Rate constant ($k_{app}$) ($min^{-1}$)** | **$R^2$** | **Degradation (%)** | **Rate constant ($k_{app}$) ($min^{-1}$)** | **$R^2$** |
| **8** | 87.9 | 0.0346 | 0.99 | 88.60 | 0.041 | 0.97 |
| **9** | 96.9 | 0.0672 | 0.96 | 92.98 | 0.049 | 0.98 |
| **10** | 99.3 | 0.0845 | 0.94 | 97.94 | 0.061 | 0.90 |
| **11** | 97.6 | 0.0619 | 0.97 | 94.86 | 0.051 | 0.99 |

## 4. Mechanism for Photocatalytic degradation:

The mechanism for photocatalytic degradation of MB in water is well established [15], [16], [44]. To analyze the degradation mechanism of MB using the as-synthesized photocatalyst, the conduction ($E_{CB}$) and valence ($E_{VB}$) band edges are calculated from the Mulliken electro-negativity theory using the following relation: $E_{CB} = \chi - E^e - 0.5E_g$ and $E_{VB} = E_{CB} + E_g$, where $\chi$ is the electronegativity of $TiO_2$ which is 5.81 eV [45], $E^e$ = free electron energy in the hydrogen scale which is 4.5 eV, and $E_g$ is the band gap energy. The calculated value of $E_{CB}$ was -0.35 eV for S1 and -0.30 eV for S2, while the $E_{VB}$ changed from +2.98 eV in S1 to +2.93 eV in S2. Generally, two degradation pathways are feasible: oxidative path, a combination of OH/$H_2O$ with $h^{+}$, and reductive path, a combination of $O_2$/$H^+$with $e^-$. The $E_{VB}$ of both samples are more positive to the redox potential of OH-/·OH (i.e., 1.9 eV) and $H_2O$/·OH radicals (i.e., 2.72 eV). Hence, from the thermodynamic point of view, h+ can react with the OH- and $H_2O$ directly to give highly reactive $\cdot OH$ radicals. Then, the reactive. Similarly, the electron from the CB combined with adsorbed $O_2$ generates $\cdot O_2^-$ radicals. The active involvement of $\cdot OH$ and $\cdot O_2^-$ radicals then convert the organic dye molecule into less harmful byproducts like $CO_2$ and $H_2O$ [11], [46], [47].

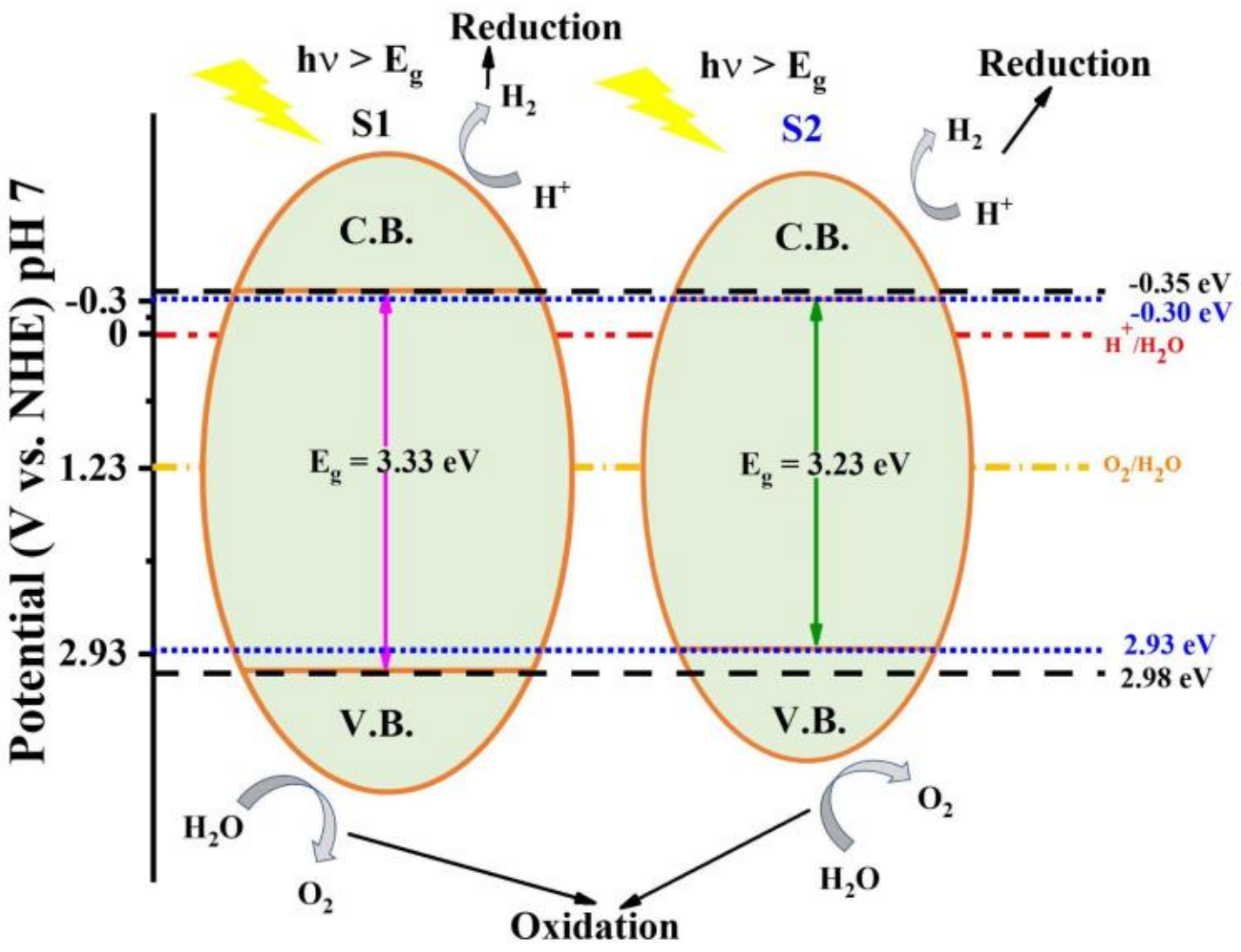

***Figure 10:*** *Schematic of generalized photocatalytic degradation mechanism for organic dyes using both S1 and S2 catalysts.*

Additionally, the efficiency of MB dye degradation may be relatively low at highly acidic pH values (1 to 3) [25]. This is because acidic conditions can hinder the generation of hydroxyl radicals (•OH), which play a crucial role in the oxidation of MB. The concentration of hydroxide ions ($OH^-$) is low, limiting the formation of •OH radicals and reducing the degradation efficiency. The efficiency of MB dye degradation tends to increase in the near-neutral to slightly alkaline pH range (4 to 10). As the pH increases, the concentration of hydroxide ions ($OH^-$) also increases, promoting the generation of •OH radicals. The higher concentration of •OH radicals enhances the oxidation of MB molecules, leading to more efficient degradation [48]. The efficiency of MB dye degradation may decrease at highly alkaline pH values (11 to 14). The high hydroxide ions ($OH^-$) concentration can lead to more competition between the MB dye and hydroxide ions for reaction with the •OH radicals. This can reduce the available •OH radicals for MB degradation, leading to a decrease in efficiency.

In this study, the high degradation efficiency observed at pH 10 can be attributed to several factors [29]:

- Enhanced hydroxyl radical generation: At pH 10, hydroxide ions (OH- ) concentration is relatively high. These $OH^-$ ions can react with photogenerated holes ($h^+$) on the photocatalyst surface, facilitating the formation of highly reactive hydroxyl radicals (•OH). Hydroxyl radicals are strong oxidizing agents and play a crucial role in the degradation of MB.
- Enhanced dye adsorption: At pH 10, the photocatalyst's surface charge may favor MB dye molecules' adsorption. The photocatalyst surface may have a positive charge at this pH, while the MB dye may carry a negative charge due to its ionization. The electrostatic attraction between the positively charged photocatalyst surface and the negatively charged MB dye molecules promotes their adsorption, increasing the contact between the dye and the photocatalyst surface. This facilitates the degradation process by providing more opportunities for the dye molecules to interact with the active sites on the photocatalyst.
- Optimal reaction conditions: pH 10 may create an environment favorable for the overall photocatalytic degradation process. It may provide the ideal balance of surface charge, adsorption capacity, and reaction kinetics for efficient degradation of MB.

Degradation efficiency decreases at higher pH (>11) values for several reasons:

- Competing reactions: Higher pH conditions can favor other chemical reactions that may compete with the photocatalytic degradation process. For example, hydroxide ions (OH-) present at higher pH can scavenge reactive oxygen species (ROS) like hydroxyl radicals (•OH) generated during photocatalysis, reducing their availability for dye degradation.

So, the enhanced photocatalytic degradation of MB dye at pH 10 can correlate well with the obtained result. Between the two samples, the higher degradation of MB dye by the S1 sample can be credited to the higher surface area, thus creating more active sites for the degradation mechanism to proceed than S2. Table 3 gives a comparative study of the proposed work with existing literature.

**Re-usability test:**

The stability and reusability of the NPs for the MB dyes were examined through five cycles of the degradation process. Following each cycle, the samples were collected via centrifugation, washed with ethanol and distilled water, and dried using a microwave oven. These dried samples were then reused for the photocatalytic dye degradation experiment, maintaining consistent experimental conditions, such as temperature and catalyst amount. Fig. 11(c, d) visually presents the reusability of the prepared nanoparticles for MB dye, demonstrating a gradual decrease in efficiency across consecutive cycles, potentially resulting from the loss of nanomaterials during the filtering and washing stages.

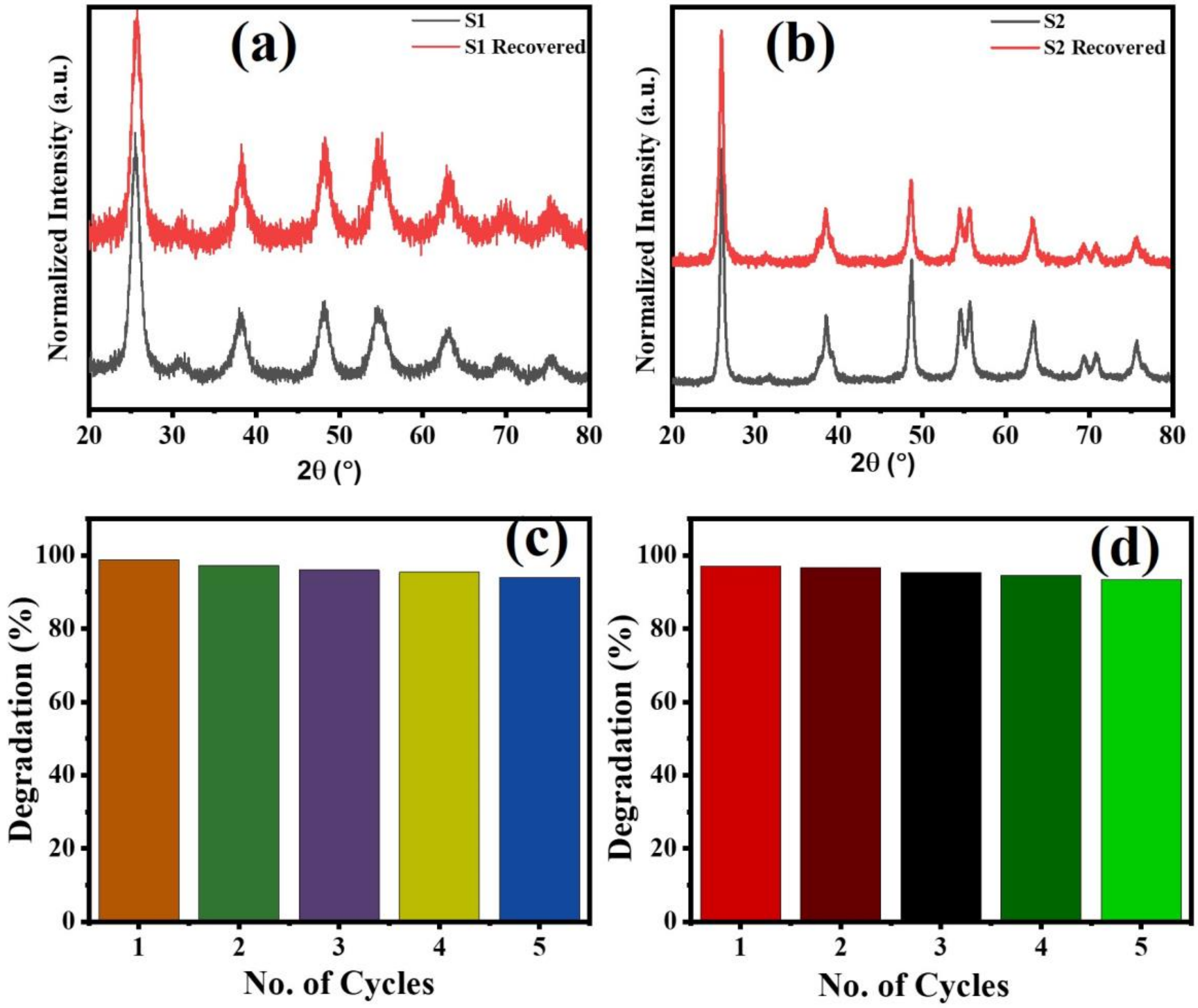

***Figure 11:*** *(a, b) XRD pattern of the fresh and recovered samples, (c, d) reusability test showing degradation percentage for five times uses of S1 and S2, respectively.*

***Table 3.*** *A comparative study of the proposed work with existing literature.*

| **Photocatalyst** | **Synthesis method** | **Photocatalytic condition Light Source, MB conc., catalyst** | **Degradation % and time** | **Rate constant ($min^{-1}$)** | **Ref.** |
|---|---|---|---|---|---|
| Mixed $MnTiO_3/TiO_2$ | Sol-gel | Sunlight, MB concentrations from $2×10^{-6}$ M to $1.5×10^{-6}$ M, Photocatalyst Conc. = 0.1 mg/mL. | 75% at 240 min. 15% to 60% increment by increasing pH from 2 to 9. | $5.96×10^{-3}$ | [49] |
| Cu doped $TiO_2$ | Sol-gel | 365 nm UV lamps MB concentration = 15 mg/l. Photocatalyst Conc. = 2 mg/mL. | Dark Adsorption = 0, 1.16, and 7.71 mg/g at pH 3, 6, and 9 Complete degradation time decreased from 300 to 30 min by increasing pH from 3 to 9. | 0.0169 at pH 3, 0.0289 at pH 6, 0.088 at pH 9 | [50] |
| $TiO_2$ | Hydrothermal | UV lamps, MB concentration = 20 mg/l, Photocatalyst Conc. = 0.05 mg/mL. | 74% at pH 7 within 120 min | 0.0101 | [51] |
| $TiO_2$-rGO powder | Hydrothermal | UV lamps<br>MB concentration = 20 mg/l.<br>Photocatalyst Conc. = 0.05 mg/mL. | 93.4% at pH 7 within 120 min | 0.0217 | [51] |
| $TiO_2$-GO aerogel | Hydrothermal | UV lamps<br>MB concentration = 20 mg/l.<br>Photocatalyst Conc. = 0.05 mg/mL. | 84.2% at pH 7 within 120 min | 0.0142 | [51] |
| $TiO_2$/reduced graphene oxide aerogel | Hydrothermal | UV lamps<br>MB concentration = 20 mg/l.<br>Photocatalyst Conc. = 0.05 mg/mL. | 99.3% at pH 7 within 120 min. | 0.0385 | [51] |
| $Fe_2O_3$(7%)-$TiO_2$ NPs | Sol-gel | UV Irradiation<br>MB concentration = 0.00001 mol/L<br>Catalyst: NM | >95%, 1 h | 0.042 | [52] |
| Pure $TiO_2$ NPs | Sol-gel | UV lamps, Visible Irradiation<br>MB concentration = 10 mg/l.<br>Photocatalyst Conc. = 0.1 mg/mL. | UV lamp, 100%, 150 min | NM | [53] |
| Co (0.24%) doped $TiO_2$ NPs | Sol-gel | UV lamps, Visible Irradiation<br>MB concentration = 10 mg/l.<br>Photocatalyst Conc. = 0.1 mg/mL. | UV lamp,<br>80 %, 150 min<br>Visible light,<br>62%, 150 min | NM | [53] |
| $TiO_2$ NPs | Hydrothermal | UV lamps<br>MB concentration = 15 mg/l.<br>Photocatalyst Conc. = 0.25 mg/mL. | 11.34% at pH = 1.4<br>38.39% at pH = 7<br>90.13% at pH = 12.6<br>Time = 1.5 h | NM | [54] |
| Polyaniline sensitized $TiO_2$ Nanocomposites (1:500) | Dispersion polymerization method | 110 W high-pressure sodium lamp,<br>MB concentration = 10 mg/l.<br>Photocatalyst Conc. = 1 mg/mL. | 81.74%,2 h | 0.01515 | [55] |
| 2wt% Ag modified $TiO_2$ | Laser-induced deposition of Ag on $TiO_2$ | 150 W<br>halogen lamp<br>MB concentration = 10 mg/l.<br>Photocatalyst Conc. = 10 mg/mL. | pH = 6.9<br>82.3%, 2 h | 0.014 | [56] |
| $SiO_2$-$TiO_2$ (2:1) NPs | Sol-gel | 8 W UV light<br>MB concentration = 10 mg/l.<br>Photocatalyst Conc. = 1 mg/mL | 85%, 30 min | NM | [57] |
| Fe-doped $TiO_2$ NPs | Hydrothermal | 300 W mercury vapor lamp<br>MB concentration = 100 mg/l.<br>Photocatalyst Conc. = 0.5 $kg/m^3$ | >90%, 65 min | NM | [58] |
| $TiO_2$/rGO (8 mass % rGO) nano composites | Hydrothermal | UV Light irradiation<br>Catalyst:  0.8 g/L<br>MB Concentration: 10 mg/L | 91.48%, 60 min | 0.0417 | [59] |
| $BiVO_4$-$TiO_2$ (1:1) nano composites | Hydrothermal | Sunlight,<br>Catalyst: 1 g/L<br>MB Concentration: $2\times10^{-5}$ M | 85%, 2h | NM | [60] |
| N-doped Graphene QD-$TiO_2$ composites | Hydrothermal | UV Light irradiation<br>MB concentration = 10 mg/l.<br>Photocatalyst Conc. = 1 mg/mL | pH=7<br>85%, 70 min | NM | [61] |
| $TiO_2$ (S1) | Hydrothermal | Sunlight irradiation MB concentration = 10 mg/l. Photocatalyst Conc. = 1 mg/mL | 87.9% at pH = 8<br>96.9% at pH = 9<br>99.3% at pH = 10<br>97.6% at pH = 11,<br>40 min | 0.035 at pH 8<br>0.068 at pH 9<br>0.084 at pH 10<br>0.062 at pH 11 | This Result |
| $TiO_2$ (S2) | Hydrothermal | Sunlight irradiation MB concentration = 10 mg/l. Photocatalyst Conc. = 1 mg/mL | 88.6 % at pH = 8<br>92.9 % at pH = 9<br>97.9 % at pH = 10<br>94.8 % at pH = 11,<br>40 min | 0.041 at pH 8<br>0.050 at pH 9<br>0.061 at pH 10<br>0.051 at pH 11 | This Result |

**Conclusion:**

The effect of precursor concentration on the size and surface morphology of $TiO_2$ nanostructures prepared by facile hydrothermal synthesis is studied. The effectiveness of the photocatalyst on MB was tested under natural sunlight. A lower concentration of the precursor, Titanium Tetra-isopropoxide, resulted in a lower pH (~ 4.6) of the precursor solution, while the higher concentration (~10 times) resulted in a higher pH (~ 5). This method of modifying the pH of the precursor solution without adding any acidic or basic reagents by slightly adding the same precursor is a unique method of altering the size of the nanostructures. The surface area could be enhanced by 3.8 times, whereas the rate constant increased by 28%. The degradation efficiency was also improved. Almost complete degradation of MB solution (10 ppm) could be achieved within 40 minutes using direct natural sunlight. Two other parameters, the concentration of photocatalyst and reaction pH, were also optimized to maximize the efficiency of the photocatalytic degradation. The optimal concentration was found to be 1 mg/mL, while the optimal pH was found to be 10. The low value of the required concentration and a low basic solution is a futuristic pathway for efficiently degrading MB dye in wastewater for large-scale industrial applications.

**Acknowledgment:**

The authors are grateful to the Department of Science and Technology (DST) of the Government of India for funding their research project (DST/TDT/AMT/2017/200). SCB thanks DST INSPIRE for providing fellowships (IF190617) for the research. MP expresses appreciation to the Ministry of Education, Government of India, for the Prime Minister Research (PMRF) and DS to UGC JRF India (211610017185) for the fellowship. FN is thankful to CSIR, India, for granting fellowships [File No-09/263(1212)/2019-EMR-I]. The authors thank the Sophisticated Instrument Centre (SIC) for access to the BET surface area analyzer facility at IIT Indore. The authors extend their appreciation to the Department of Science and Technology (DST), Govt. of India, for allocating a FIST instrumentation fund to the discipline of physics at IIT Indore to purchase a Raman Spectrometer (Grant Number SR/FST/PSI-225/2016). The authors thank Dr. Jaydeep Bhattacharya from Jawaharlal Nehru University, India, for his assistance with DLS measurements.